\documentclass{article}

    \PassOptionsToPackage{numbers, compress}{natbib}

    \usepackage[final]{neurips_2023}

\usepackage[utf8]{inputenc} 
\usepackage[T1]{fontenc}    
\usepackage{hyperref}       
\usepackage{url}            
\usepackage{booktabs}       
\usepackage{amsfonts}       
\usepackage{nicefrac}       
\usepackage{microtype}      
\usepackage{xcolor}         

\usepackage{graphicx}	
\usepackage{amsmath}	
\usepackage{amssymb}	
\usepackage{bm}
\usepackage[percent]{overpic}
\usepackage[normalem]{ulem}
\usepackage{tikz}
\usetikzlibrary{arrows,positioning}
\usepackage[font=small,labelfont=bf]{caption}
\usepackage{makecell}

\usepackage{caption}
\usepackage{subcaption}

\title{Predicting the Age of Astronomical Transients from Real-Time Multivariate Time Series}

\author{%
Hali Huang$^{1}$ \quad  Daniel Muthukrishna$^{1*}$ \quad Prajna Nair$^1$ \quad Zimi Zhang$^1$  \\
 \textbf{Michael Fausnaugh}$^{1,2}$ \quad \textbf{Torsha Majumder}$^1$ \quad \textbf{Ryan J. Foley}$^3$ \quad \textbf{George R. Ricker}$^1$  \\
$^1$MIT \quad $^2$Texas Tech University  \quad $^3$UC Santa Cruz\\
\texttt{danmuth@mit.edu}\\
}

\begin{document}

\maketitle

\begin{abstract}
Astronomical transients, such as supernovae and other rare stellar explosions, have been instrumental in some of the most significant discoveries in astronomy. New astronomical sky surveys will soon record unprecedented numbers of transients as sparsely and irregularly sampled multivariate time series. To improve our understanding of the physical mechanisms of transients and their progenitor systems, early-time measurements are necessary. Prioritizing the follow-up of transients based on their age along with their class is crucial for new surveys. To meet this demand, we present the first method of predicting the age of transients in real-time from multi-wavelength time-series observations. We build a Bayesian probabilistic recurrent neural network. Our method can accurately predict the age of a transient with robust uncertainties as soon as it is initially triggered by a survey telescope. This work will be essential for the advancement of our understanding of the numerous young transients being detected by ongoing and upcoming astronomical surveys. 

\end{abstract}

\section{Introduction}
\label{sec:Introduction}


The study of astronomical transients is crucial for understanding the nature of the universe, from one class of supernovae being a fundamental probe for measuring the expansion rate of the universe, to probing nucleosynthesis in extreme conditions. New astronomical sky surveys such as the Vera Rubin Observatory's Legacy Survey of Space and Time (LSST, \citealt{Ivezic2009LSST:Products}) are expected to observe over 10 million transient alerts each night, increasing the discovery rate of known supernovae by 100 times. Given these massive data volumes and the limited number of worldwide follow-up observatories, only a tiny fraction of these events can expect to receive detailed spectroscopic or multi-wavelength follow-up observations. 

In recent years, several efforts have been made to automate the identification of transients using machine learning techniques. However, most previous approaches require the full phase coverage of the transient, limiting the scientific questions that can be answered about these events. Early-time follow-up observations, including high-cadence photometry and time-resolved spectroscopy shortly after a transient's explosion, are necessary for gaining insights into the progenitor systems driving the event and advancing our understanding of the transient's physical mechanisms \citep{Kasen2010}. This has motivated the need for the early identification of young transients (such as the Young Supernova Experiment \citep{Jones2021-YSE}), which is becoming increasingly critical with the volume of data generated by new survey telescopes \citep[e.g.][]{Muthukrishna19RAPID, Muthukrishna2022,Villar2021_Anomalydetection, Pimentel2023, Allam2023}. However, early classification on real data has not yet been effectively demonstrated in the literature. Therefore, we propose that accurately measuring a transient's age is a crucial criterion for selecting which transients to follow up from the abundance of new data. In this paper, we present the first method for real-time identification of a transient's age, enabling the prioritization of young transients and facilitating rapid follow-up and detailed observations of their progenitor systems. Using this framework alongside transient classifiers and anomaly detection frameworks will be essential for making discoveries in new large-scale surveys.



\section{Data}
To evaluate our method, we collected a set of labeled transients that have been observed by the Zwicky Transient Facility (ZTF, \citep{Bellm2019ZTF}) and the Transiting Exoplanet Survey Satellite (TESS, \citep{TESS_Ricker_2015}). Our dataset comprises 805 transients, each represented by a multi-channel time series known as a \textit{light curve}. 

The majority of transients, 529 of them, in our dataset are unclassified. The classified transients consist of 230 Type Ia Supernovae, 37 core collapse supernovae and 9 other types. Type Ia supernovae result from the thermonuclear explosion of a binary system with a carbon-oxygen white dwarf accreting matter from a companion star. Core collapse supernovae include Type Ibc (SNIbc) and Type II (SNII) supernovae and  are the result of the core collapse of massive stars. The other category consists of a variety of classifications found uncommonly in the dataset such as Tidal Disruption Events (TDE), variable stars (CVs), and Type I Superluminous supernovae (SLSN).

These light curves include flux and flux uncertainty measurements in the ZTF-$g$ (green wavelength range), ZTF-$r$ (red wavelength range), and TESS (red to infrared wavelength range) passbands. Notably, the light curves are sparsely and irregularly sampled. The ZTF passbands exhibit a mean cadence of 3-day intervals, while the TESS bands have a mean cadence of 10 minutes but cover only a short duration ($\sim 27$ days) of the light curve. An example light curve is depicted in Fig.~\ref{fig:example_lc}.



We define the \textit{trigger} as the time when ZTF first detects a $5\sigma$ change in the flux. We exclude observations more than 30 days before the trigger and more than 70 days after the trigger to focus on the transient phase of the light curves. We used the forced photometry service from ZTF \citep{Masci2023-ZTF-forced-photometry} to quantify the non-detection observations early in a  transient light curve. These early measurements contain crucial information about the host galaxy that has been shown to be very useful for transient identification and modeling \citep[e.g.][]{Foley2013ClassifyingData, Gagliano2021}. 





\section{Method}
The goal of this work is to predict the age of a transient based on photometric time-series data. The age of a transient is typically defined by either the number of days after maximum light or the number of days after first light, which is approximately coincident with the supernova explosion time. We have built separate models to predict the time of first light $t_0$ and the time of maximum light $t_{\mathrm{max}}$. It is important that the age can be identified as early as possible - ideally within a few days after the time of trigger.

Dealing with missing data is a major challenge in time-series machine learning, with most works opting for some sort of interpolation of the data. However, these interpolation methods can introduce out-of-distribution data sets that are not representative of real data. In our work, we bypass the missing data problem by explicitly providing the neural network with time $t$ relative to trigger for each data point. We set each column of the input matrix representing the multi-channel time-series data set as follows,
\begin{equation}
     \bm{X}_{spt} = [t, \lambda_p, D_{spt}, \sigma_{D,spt}]^\top,
\end{equation}
where $D_{spt}$ is the observed flux for transient $s$, passband $p$, at time $t$, $ \sigma_{D,spt}$ is the corresponding flux uncertainty, and $\lambda_p$ is the central wavelength of each passband in units of $10^{-7} \mathrm{m}$ as follows \{$\lambda_g=4.767$, $\lambda_r=6.215$, $\lambda_\mathrm{TESS}= 7.865$\}. Hence, each column of the matrix represents an observation with the flux and uncertainty at a particular time and passband. 

\subsection{Obtaining Ground Truth Labels}
\label{sec:ground_truth_labels}
In order to apply supervised learning methods to our dataset, we require ground truth labels for the ages of each transient. Because of missing light curve data, the ground truth is unknown, and we need to estimate the times of maximum and first light. We developed a Bayesian parametric model to estimate the times and their uncertainties.
The training labels for the time of maximum light were calculated by fitting the full light curves with the Supernova Parametric Model (SPM \citep{Sanchez-Saez2021_SPM}), which is as follows:
\begin{equation}
F=\frac{A(1-\beta'(\frac{t-t_0}{t_1-t_0}))}{1+e^{-\frac{t-t_0}{\tau_{\mathrm{rise}}}}}.\left[1-\sigma \left(\frac{t-t_1}{3} \right) \right]+\frac{A(1-\beta')e^{-\frac{t-t_1}{\tau_{\mathrm{fall}}}}}{1+e^-{\frac{t-t_0}{\tau_{\mathrm{rise}}}}}. \left [\sigma \left(\frac{t-t_1}{3}\right) \right]
\end{equation}
where $F$ is the model flux of the light curve and $A$, $t_0$, $t_1$, $\tau_{\mathrm{fall}}$, $\tau_{\mathrm{rise}}$, $\beta'$ are free parameters that were tuned for each light curve. We set uniform priors for each of the parameters to constrain them to physically realistic values. We then drew sample parameters of the posterior function using Markov Chain Monte Carlo (MCMC) sampling. We obtained the times of maximum light $t_{\mathrm{max}}$ along with uncertainties $\sigma_{t_{\mathrm{max}}}$ from the posterior predictive light curves. 

We similarly obtained the times of first light by fitting the early light curve at times $t<t_{\mathrm{max}}$ with a power law model as defined in equation 1 from \citet{Vallely2021} as follows,
\begin{equation}
F= H(t-t_0) . C. (t-t_0)^{\beta_1(1+\beta_2(t-t0))} + f_0,
\end{equation}
where $C$, $t_0$, $f_0$, $\beta_1$, and $\beta_2$ are free parameters that were tuned for each light curve, $t_0$ is the time of first light, and $H$ is the Heaviside step function. We set uniform priors to keep the free parameters within physically realistic ranges and then used MCMC sampling to obtain posterior samples of $t_0$. We used the mean and standard deviation of these samples as the target labels of the time of first light $t_0$ and corresponding uncertainty $\sigma_{t_0}$, respectively. 
We use both the mean and uncertainties of the target ages in the loss function of the neural network model.

\subsection{Probabilistic Recurrent Neural Network}
We developed a probabilistic recurrent neural network (RNN) built using Gated Recurrent Units (GRUs, \citep{GRU}) to map an input multi-passband light curve matrix, $\bm{X}_{s(t \le T)}$, for transient $s$ up to time $T$ onto a probabilistic prediction of the transient age,
\begin{equation}
    \bm{Y}^w_s = \bm{f}(\bm{X}_{s(t \le T)} ; \bm{w}),
\end{equation}
where $\bm{w}$ are the parameters (i.e. weights and biases) of the network, and $\bm{Y}^w_s$ has been parameterized as a Normal distribution with a predicted mean age $\hat{y}_s^w$ and a predicted uncertainty $\hat{\sigma}^w_s$ for a particular set of network weights $w$. We have built two distinct RNN models: one to predict the time of first light with a predicted mean $\hat{y}_s^w=\hat{t}_0^w$ and uncertainty $\hat{\sigma}_s^w = \hat{\sigma}_{t_0}^w$ and another to predict the time of maximum light with a predicted mean $\hat{y}_s^w=\hat{t}_{\mathrm{max}}^w$ and uncertainty $\hat{\sigma}_s^w = \hat{\sigma}_{t_{\mathrm{max}}}^w$. Each model is designed to 
regress over past data to make real-time predictions that get updated as new data are observed. The model architectures and inputs are identical, with the only difference being the output target variable.


The recurrent neural network architecture comprises two GRU layers, each consisting of 50 units with ReLU activation functions. These GRU layers are unidirectional, meaning only information from past timesteps are used for prediction. This is to ensure that the network makes temporally causal real-time predictions that update as additional data becomes available. The final layer of the network is a time-distributed fully-connected probabilistic layer. It connects all 50 neurons from the second GRU layer to an output vector $\bm{Y}_s^w$ of size 2 at each timestep for the predictive mean and standard deviation. Each layer in the network includes batch normalization and dropout regularization with a dropout rate of 20\%. The input to the network is the matrix $\bm{X}_s$. 


\subsubsection{Uncertainty Quantification}
Applications of deep learning to astronomy have often avoided uncertainty quantification in favor of point estimation. Here, we exemplify a simple methodology for quantifying uncertainty for any NN regression problem. 

To characterize the intrinsic uncertainty of our network's ability to predict age, we build a probabilistic network and parameterize the output as a Normal distribution. Therefore, instead of the typical neural network that only predicts a point estimate, our network outputs a predictive mean $\hat{y}_s^w$ and a standard deviation $\hat{\sigma}^w_s$ for a particular set of weights $\bm{w}$. The uncertainty $\hat{\sigma}^w_s$ is helpful to model the intrinsic error of our model, such that even if we had no measurement error and an infinite training set, there would still be some discrepancy between our model predictions and the true age of the transient. 

We then also quantify the epistemic uncertainty of our model's predictions of $\bm{Y}^w_s$ by making an approximate Bayesian neural network. We employ an approach by \citet{Gal2015} called Monte Carlo (MC) dropout sampling, which has been shown to provide posterior samples of the network parameters when the commonly used dropout regularization technique is applied during inference instead of just during training. We collect the results of stochastic forward passes through the network as approximate posterior draws of $\bm{Y}^w_s$ and use the mean and standard deviation of these draws as our marginal predictive mean $\hat{y}_s$ and predictive uncertainty $\hat{\sigma}_s$.

The predictive distribution for the underlying latent age $y_s(\bm{w})$ of a transient is 
\begin{equation}
      \mathcal{P}(y_s|\bm{X}_{s(t \le T)}, \bm{w}) = \mathcal{N} \left( y_s(\bm{w}) \mid  \hat{y}_s^w, \hat{\sigma}^w_s \right),
\label{eq:posterior_predictive}
\end{equation}
where the latent age $y_s(\bm{w}) = \hat{y}_s^w +  \epsilon_{s}^w$ is modeled by the neural network predictive mean $\hat{y}_s^w$ with an associated error $\epsilon_{s}^w \sim \mathcal{N}\left(0, \hat{\sigma}^w_s\right)$ that is a zero-mean Gaussian random variable with standard deviation $\hat{\sigma}^w_s$.


Next, a generative model of the observed age $\gamma_s$ is derived by adding a measurement error to the latent age, $\gamma_s = \hat{y}_s^w + \epsilon_{s}^w + \zeta_{s}$, where we assume that the measurement error of the age $\zeta_{s} \sim \mathcal{N}\left(0, \eta_s\right)$ is a zero-mean Gaussian random variable with standard deviation $\eta_s$. Then, we define the cost function of the neural network as the likelihood function,
\begin{equation}
      \mathcal{P}(\gamma_s|\bm{X}_{s(t \le T)}, \bm{w}) = \mathcal{N} \left( \gamma_s \mid \hat{y}_s^w, \sqrt{\hat{\sigma}^w_s + \eta_s}  \right)
\label{eq:DNN_likelihood}
\end{equation}
where we include both predicted and measured uncertainties $\eta_s$ in quadrature. The Bayesian posterior distribution over the weights is 
\begin{equation}
     \mathcal{P}(\bm{w} |\bm{X}) \propto \mathcal{P}(\bm{w}) \prod\limits_{s}\prod\limits_{T} \mathcal{P}(\gamma_s|\bm{X}_{s(t \le T)}, \bm{w}),
\label{eq:DNN_posterior}
\end{equation}
where we take the product of the likelihood over all times and all transients in the training set, and the prior is defined as a zero-mean Normal distribution. 
Finally, to make a prediction of the age of a transient, we evaluate the posterior predictive distribution of the age as follows,
\begin{equation}
    \mathcal{P}(y_s | \bm{X}_{s(t \le T)}) = \int \mathcal{P}(y_s | \bm{X}_{s(t \le T)}, \bm{w}) \mathcal{P}(\bm{w} |\bm{X}) d\bm{w},
    \label{eq:DNN_predicitve_distribution}
\end{equation}
where we marginalize over the weights of the network by integrating the product of the predictive distribution of the age (first term in the integrand, defined in equation \ref{eq:posterior_predictive}) and the posterior distribution over the network weights (second term in the integrand, defined in equation \ref{eq:DNN_posterior}). The integral is intractable, so we approximate it using MC dropout \citep{Gal2015}. Each inference produces a sample of the posterior; thus, we draw 100 samples $i$ each with a predicted age $y_{s, i}(\bm{w}_i) \sim \mathcal{N}(\hat{y}_s^w, \hat{\sigma}^w_s)$ by executing 100 forward passes of the neural network for a given input. We compute the marginal predictive mean $\hat{y}_s$ and predictive uncertainty $\hat{\sigma}_s$ as the sample mean and standard deviation of the 100 samples, and report this for each transient.






\begin{figure}
    \centering
    \begin{subfigure}[b]{0.495\textwidth}
         \centering
         \includegraphics[width=\textwidth]{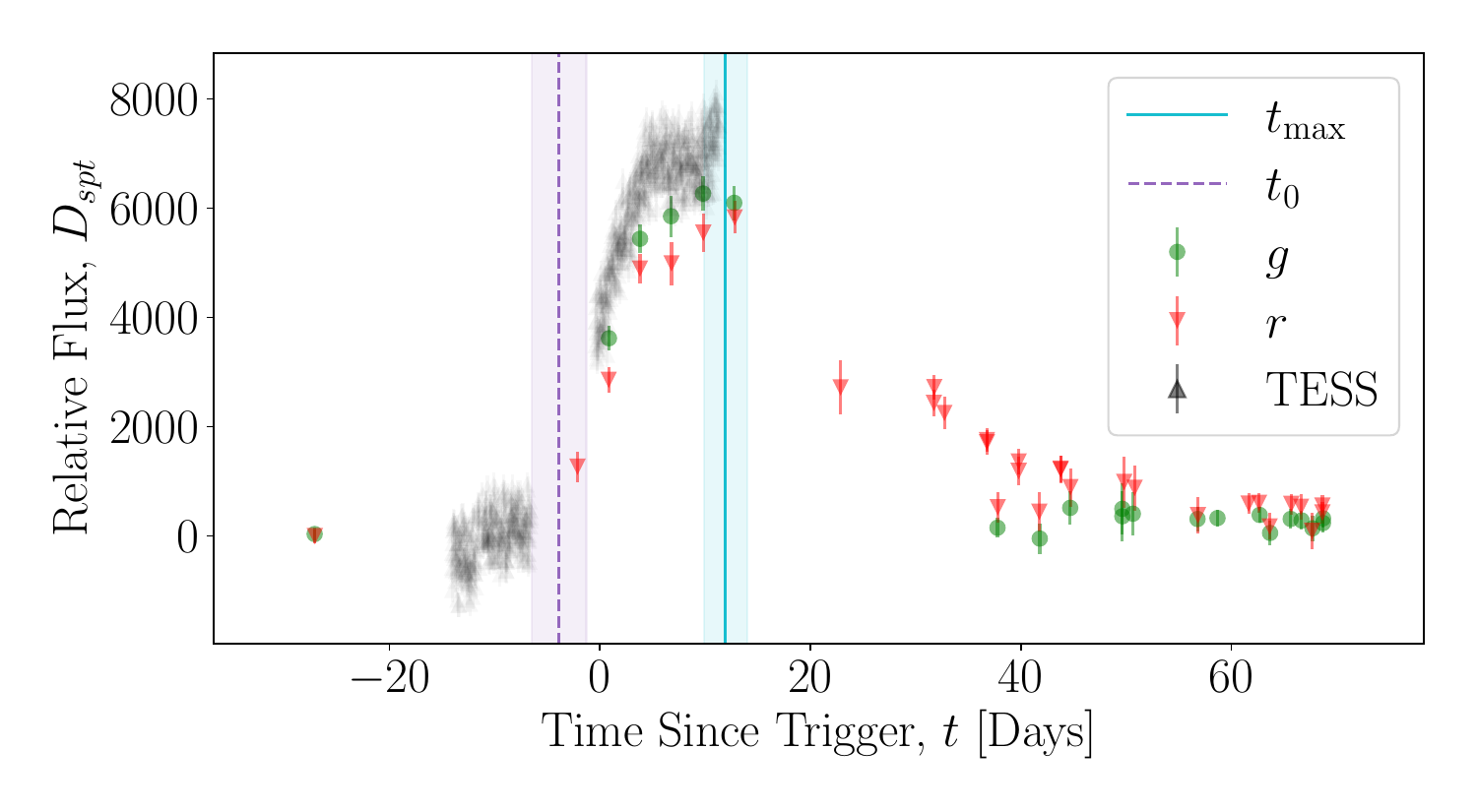}
         \vspace{-1.5em}
         \caption{}
         \label{fig:example_lc}
    \end{subfigure}
    \hfill
    \begin{subfigure}[b]{0.495\textwidth}
         \centering
         \includegraphics[width=\textwidth]{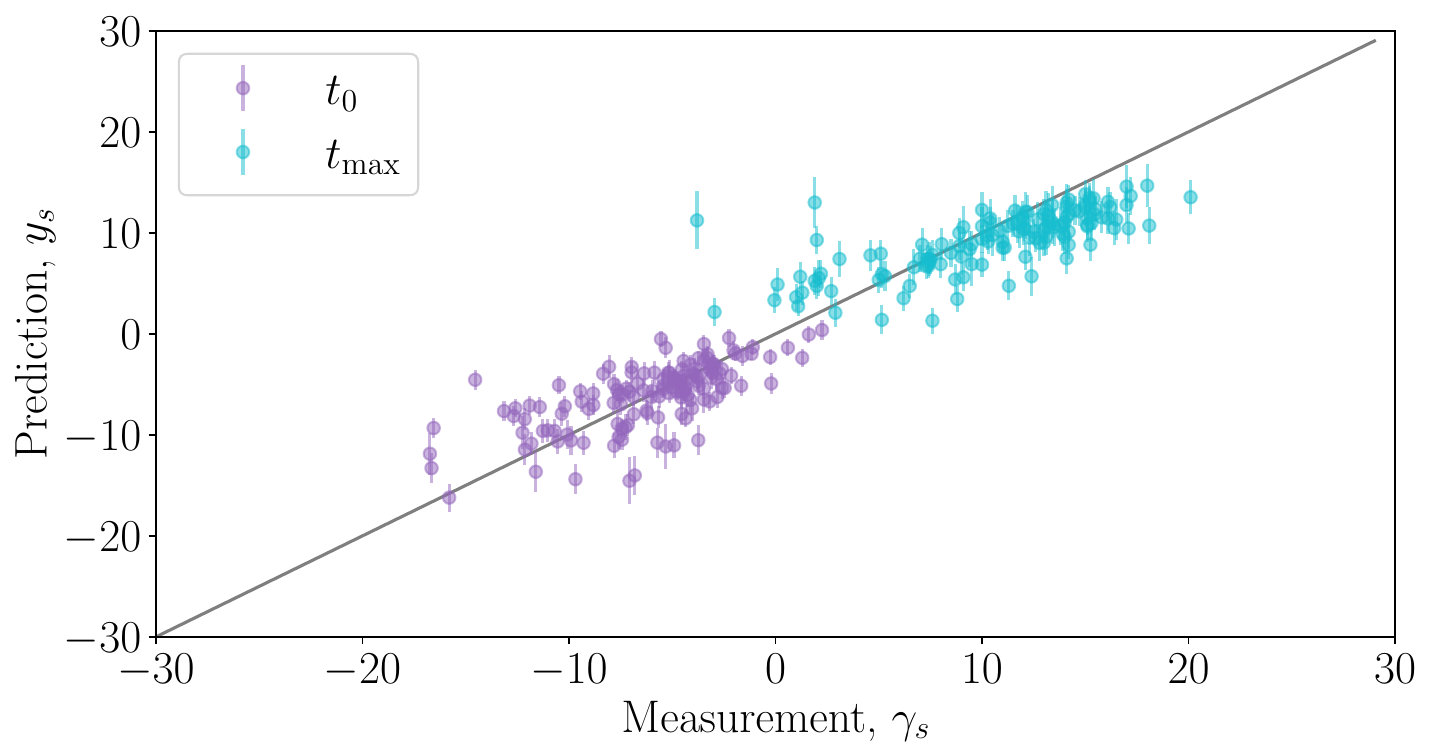}
         \vspace{-1.5em}
         \caption{}
        \label{fig:pred_vs_actual}
    \end{subfigure}
    \caption{(\textbf{a}) An example light curve of a transient (SN2018hyy) in our dataset. The vertical cyan and dashed purple lines represent the predicted time of maximum light $t_\mathrm{max}$ and explosion light $t_0$, respectively, with the shaded bar illustrating the predicted uncertainties $\sigma_s$. 
    (\textbf{b)} A comparison of the predicted and measured ages of all transients in the test set given partial light curve data at 20 days after trigger. 
    }
    \vspace{-1em}
\end{figure}

\begin{figure}
    \centering
    \begin{subfigure}[b]{0.495\textwidth}
         \centering
         \includegraphics[width=\textwidth]{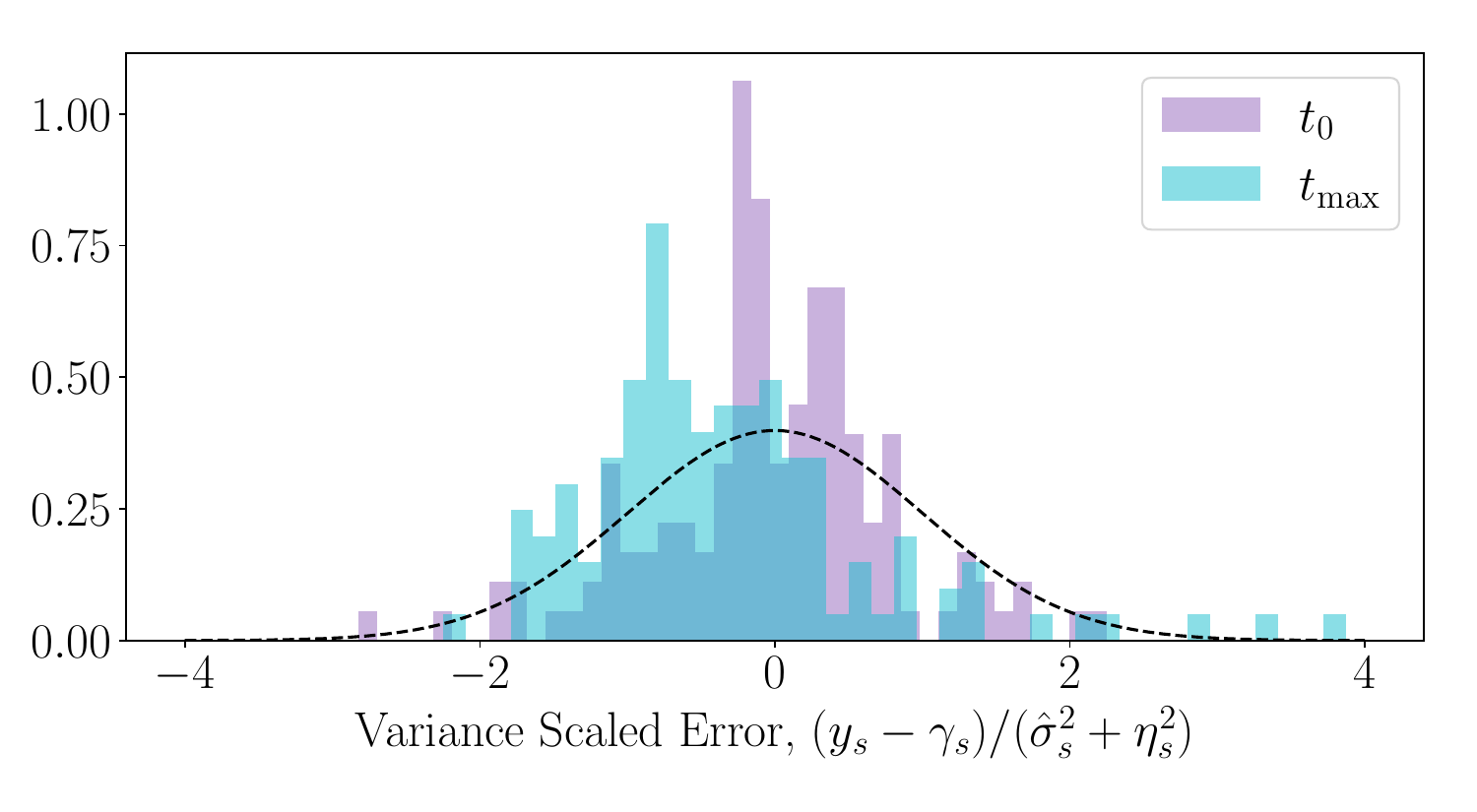}
         \vspace{-1.5em}
         \caption{}
         \label{fig:residuals}
    \end{subfigure}
    \hfill
    \begin{subfigure}[b]{0.495\textwidth}
         \centering
         \includegraphics[width=\textwidth]{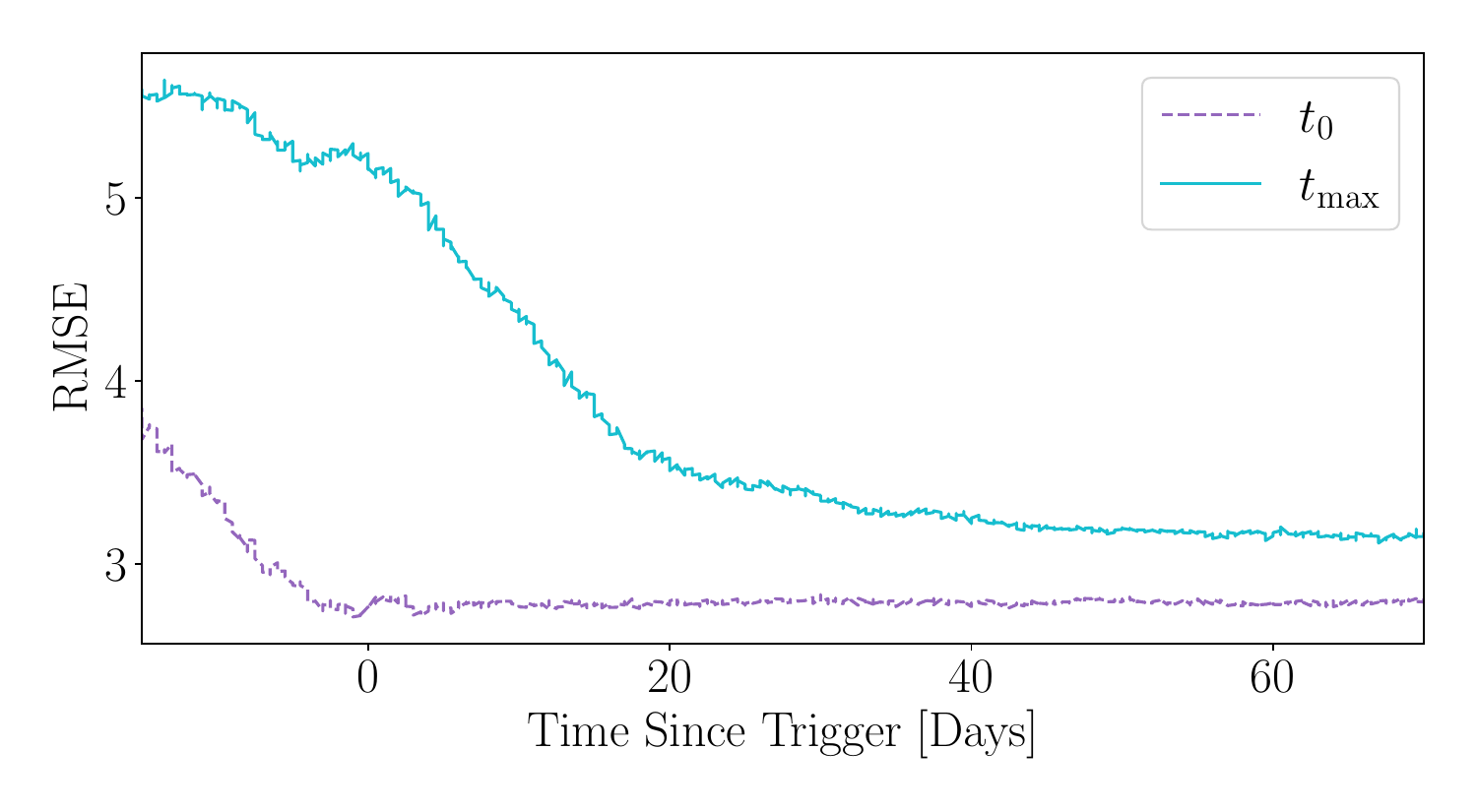}
         \vspace{-1.5em}
         \caption{}
        \label{fig:RMSE}
    \end{subfigure}
    \caption{(\textbf{a}) The distribution of the error across all time steps scaled by the predictive and measured variances. A unit Gaussian $\mathcal{N}(0,1)$ is shown as the black dashed line to evaluate the effectiveness of the predicted uncertainties. The standard deviations of the scaled errors are 0.99 and 0.81 for the $t_{\mathrm{max}}$ and $t_0$ models, respectively, and the means are both very close to zero.
    (\textbf{b)} A plot of the root-mean-square error (RMSE) between the predicted and measured values of the ages. The RNN makes a prediction at every time-step given light curve data up to time $T$, and we plot the RMSE at each time step. The plot indicates that the predictions improve as more data along the light curves become available.}
    \vspace{-1em}
\end{figure}

\section{Results and Discussion}
We have developed two distinct RNN models, one to predict the time of first light $t_0$ and the other to predict the time of maximum light $t_{\mathrm{max}}$. We trained the neural network on $\sim80\%$ of the initial dataset (663 transients), and then validated the performance of the models on the remaining 142 transients ($\sim20\%$ of the dataset). We drew 100 posterior sample predictions of $\bm{Y}_s^w$ from our network and plotted the mean and standard deviation of our predictions at 20 days after trigger against the measurements in Fig.~\ref{fig:pred_vs_actual}. The predictions closely align with the measurements, as evidenced by the scaled errors depicted in Fig.~ \ref{fig:residuals} which are distributed near zero. At the time of trigger, the mean absolute error is $\mathrm{MAE_{t_0}}=2.1$ days and $\mathrm{MAE_{t_\mathrm{max}}}=4.2$ days for the two models. Considering that the average cadence for the ZTF observations is 3 days, our algorithm impressively predicts the correct age (since explosion time) within less than the time between ZTF observations for 68\% of transients and within 6 days for 95\% of transients.

We examine how our methods perform at early times, and how they improve as new data along the light curves become available. We emphasize that, unlike previous work which predicts the time of maximum retrospectively, our work is the first to show results as a function of time and to achieve successful predictions at early times. In Fig.~\ref{fig:RMSE}, we plot the root-mean-square error (RMSE) at every time-step outputted by our RNN models. The predictions of the time of first light improve rapidly as more light curve data is recorded by the model, and plateaus at the time of trigger. Thus, as soon as a transient is first detected by ZTF, we can accurately determine the age with an RMSE less than 2.8 days. Predictions of the time of maximum have a slightly larger RMSE, and as expected, improve more after maximum light, which occurs for most transients within 20 days after trigger. For a baseline comparison, we compared our results with the simpler parametric models described in section \ref{sec:ground_truth_labels}. The parametric models had poor performance on partial light curve data and could not compete with the speed of inference of our RNN models.

Overall, this paper presents the first method for identifying the age of a transient in real-time. It is scalable to surveys as large as LSST, with an inference time of 1.7 milliseconds per execution and 0.17 seconds per object for recording 100 posterior samples when run on a single NVIDIA T4 GPU. In future work, we plan to implement this algorithm into the real-time Young Supernova Experiment survey (YSE, \citep{Jones2021-YSE}). Further work could also investigate utilizing transformers instead of RNNs for the model architecture. We briefly experimented with transformers during this research but found no improvement in accuracy. The limited size of our dataset may be a contributing factor, as a larger dataset could take better advantage of the training speed of transformers. We have evaluated our method on a set of real transient events from the Zwicky Transient Facility (ZTF) and the Transiting Exoplanet Survey Satellite (TESS), and demonstrated the effectiveness of our approach in identifying young transients with high accuracy and quantifiable uncertainties. This work, in combination with real-time transient classifiers, will be critical for discovery and enabling a deeper understanding of progenitor systems in the new era of large-scale astronomical surveys.

\begin{ack}
This paper includes data collected by the Transiting Exoplanet Survey Satellite (TESS) mission. The TESS mission is
funded by NASA’s Science Mission Directorate. The UCSC team is supported in part by NASA grant 80NSSC20K0953, NSF grant AST–1815935, and by a fellowship from the David and Lucile Packard Foundation to R.J.F.

\end{ack}

\bibliographystyle{mnras}
\bibliography{references.bib}


\end{document}